\documentstyle[preprint,aps,tighten]{revtex}
\begin{document}

\preprint{\vbox {\hspace*{\fill} DOE/ER/40762-058\\
          \hspace*{\fill} U. of MD PP\#95-118}}

\title{Baryon Isovector Electric Properties
 and the Large {\boldmath $N_c$} and Chiral Limits}

\author{Thomas D. Cohen\footnote{On leave from Department of
Physics, University of~Maryland, College~Park, MD~20742.}\\}

\address{Department of Physics, FM-15, and Institute for~Nuclear
Theory, NK-12\\
University of Washington, Seattle, WA~98195, USA}

\maketitle

\begin{abstract}
A model independent calculation is given for
 the nucleon isovector electric charge radius
 which is valid in the limit $N_c \rightarrow
 \infty$, $m_\pi \rightarrow 0$, $N_c m_\pi$ fixed.  This expression
reduces
 to that of the Skyrme model in the limit
 $N_c m_{\pi} \rightarrow \infty$.
\end{abstract}

\newpage

The study of baryons in the large $N_c$ and
 chiral limits is an interesting
and subtle problem.  The interest stems
largely from the belief that for many purposes $N_c=3$ is large
enough so that an
 expansion of physical quantities in a $1/N_c$
 expansion gives at least a  reasonable, if
 crude, description of hadronic properties
 while at the same time the up and down
 quark masses are light enough so that an
 expansion in terms of $m_q$ is also
 reasonable.  However,
for some important physical quantities
the limits $m_q \rightarrow 0$ and $N_c
 \rightarrow \infty$
do not commute \cite{an,cb}. That is, if
 such
 a quantity is treated as a function of
 $m_q$, then the $1/N_c$ expansion is not
 uniformly convergent.
  This  implies that for such
 quantities expansions in $1/N_c$
and in $m_q$ cannot be consistent with
each other.
This issue is significant for both theoretical and practical
reasons.  On
the theoretical side it helps make manifest
 the nature of QCD in these important
 limits.  As a practical matter,
if the two expansions are inconsistent, at
 least one of the expansions must be   invalid.  Thus, one may
learn about places
 where one expects the expansions to fail
 and,
 perhaps, one may be able to correct for the
 failure of one of the expansions.

This letter will explore
the large $N_c$ and chiral behavior
of isovector electric properties of baryons.
  It has long been suspected that
some of these properties are afflicted with
 problems associated with noncommuting
 limits.  The evidence for this has been
in the Skyrme model\cite{skyrme} in which
 Adkins and Nappi\cite{an} found that the
 behavior of the isovector charge radius of
 the proton
in the limit $m_{\pi}^2 \rightarrow 0$ goes
as $ m_\pi^{-1}$ while in conventional
 chiral perturbation theory one finds the
 same quantity goes as $ {\rm Log}( m_\pi)$
 times a coefficient proportional to
 $N_c$\cite{chi1,chi2}.  They speculated
 that this was due to the lack of
 commutativity of the two limits.
Here it will be shown that that this
 speculation is correct and the underlying
nature of this behavior will be explained
in terms of the role played by the
 $\Delta$ isobar.

More specifically, it will be shown: (i)
that the leading chiral behavior of the
 isovector charge radius is calculable in
 large $N_c$ chiral perturbation theory---a
 generalization of usual chiral perturbation
 which is valid in the limit
 $m_{\pi} \rightarrow 0 \; , \;  N_c
 \rightarrow \infty \; , \;  N_c m_\pi
 \rightarrow {\rm fixed}$. It is given by
\begin{equation}
\langle {\rm N}| r^2 |{\rm N} \rangle_{I=1}
 \, = \frac{5 \, g_A^2 }{4 \, \pi^2 f_\pi^2}
 \frac{\delta m}{\sqrt{m_\pi^2 - \delta m^2}} {\rm
tan}^{-1}\left(\sqrt{\frac{m_{\pi} - \delta
 m}{m_\pi + \delta m}}\right ) + C \;
\label{r2e} \end{equation}
 where $\delta m = M_\Delta - M_N$ and $C$
 is a constant subleading in either $1/N_c$,
 $m_{\pi}^2$ or both.
(ii)  There is a model-independent prediction
 of the isovector charge radius for any
 Skyrme model or other large $N_c$ hedgehog
 model, in terms of  physical observables,
 which is valid in the limit $m_{\pi}
 \rightarrow 0 \; , \; N_c \rightarrow
 \infty$ with the large $N_c$ limit  taken  first.  It is given
by:
\begin{equation}
\langle {\rm N}| r^2 |{\rm N} \rangle_{I=1}
 \, = \, \frac{5 \, g_A^2 \,  \delta m}{16 \, \pi \, f_\pi^2 \,
m_\pi}
\label{r2s} \end{equation}
(iii) This model-independent Skyrme result
 differs from that obtained in usual chiral
 perturbation theory with the limit
$m_{\pi} \rightarrow 0$ taken first.  (iv)  However, the
model-independent prediction from all
 Skyrme-type models is equivalent with the
large $N_c$ chiral perturbation theory
 prediction provided the large $N_c$ limit
is taken prior to the the chiral limit.
This provides support for the conjecture
 that the Skyrme model correctly reproduces
 all model-independent predictions of large $N_c$ QCD.
(v)The $1/N_c$ expansion for the  isovector
 charge radius may have a finite radius of
 convergence.  There is significant evidence
 that $N_c=3$ is outside this radius of
 convergence.
(vi) Analogous results can be obtained for
 other isovector electromagnetic observables
 such as the
higher moments of the electric form factor of the nucleon or the
E2 N-$\Delta$ quadrupole transition.

The $\Delta - N$ mass splitting plays a central role in
eqs.~(\ref{r2e}) and (\ref{r2s}).  This is not unexpected.  As
 has been noted previously\cite{cb}, the
 fact that
\begin{equation}
M_\Delta - M_N \sim 1/N_c
\end{equation}
can easily lead to noncommutativity
of the large $N_c$ and chiral limits. The
 reason for this clear. The leading
 nonanalytic behavior (in $m_q$ or $m_{\pi}^2$ ) for
 the charge radius in chiral perturbation
 theory  is due to  small energy
 denominators coming from nucleon plus one
 pion states.
However, in the $N_c \rightarrow \infty $
 limit, $M_\Delta - M_N \rightarrow 0$ and
 there are new states with small energy
 denominators, such as $\Delta$ plus one
 pion states which can contribute to the nonanalytic behavior.

The Skyrme model\cite{skyrme} and other
 hedgehog soliton models such as the
 chiral-quark meson soliton model\cite{cqm},
 the chiral or hybrid bag model
 \cite{hybrid:rev} or the soliton approach
 to the Nambu--Jona-Lasinio (NJL)  model
\cite{NJL} are based on both large $N_c$ and
 approximate chiral symmetry.  As all of
 these models behave identically
for all of the issues discussed here, the
phrase ``Skyrme'' model will be used in this
 letter to denote any of these hedgehog
 models.
 Approximate chiral
symmetry is explicit in the form of the
 effective lagrangian.  The large $N_c$
 nature
is built into the method of calculation.  These models are
quantum field theories which in general are computationally
intractable.  However, in the large $N_c$
 limit, the models are calculable.  The large $N_c$ limit is
essential in two ways.
 The first  is that in this limit the
 problem reduces
to a tractable classical field theory.
 Unfortunately, the classical solutions are
 hedgehog configurations which break symmetries of the theory; to
obtain physical
 results one must project onto states with
 physical quantum numbers.  There is a
  semiclassical projection method based on
 the separation of collective and intrinsic
 variables \cite{sc}. However, this
 technique is valid only in the large $N_c$
 limit.  Thus, calculations in the Skyrme
 model  implicitly correspond to working in
 the $N_c \rightarrow 0$ limit---{\it i.e.}
 to leading order in the $1/N_c$ expansion.
  The large $N_c$ character of the model is
 also manifest in that the calculated baryon
 properties will be consistent with Witten's
 large $N_c$ scaling rules for
 baryons\cite{witten:nc}.

It should be stressed that  the
 semiclassical treatment of the problem is
 in an essential ingredient of the Skyrme-type
models---unless
one specifies a
 calculational scheme,  these
 nonrenormalizable models are not well
 defined.
The fact that the large $N_c$ approximation
 is implicit in the semiclassical treatment
 used in all calculations implies that if
 some quantity has noncommuting large $N_c$
 and chiral limits, then one expects that
 semiclassical calculations in Skyrme-type
 models to correspond to taking the $N_c
 \rightarrow \infty$ first.

The value of the electric charge radius
may be calculated in a straightforward way
 in Skyrme models using semiclassical
techniques \cite{sc}.  The result depends on
 the details of the model.  However, it is
 easy to see that as $m_\pi \rightarrow 0$
the integrals are dominated by the
 contributions from the tail of the
 distribution. The pion fields in a hedgehog
 configuration may be written as
$\pi_a(\vec{x}) = \pi(r)\hat{r}_a$ where $a$
 is the isospin direction.  In any Skyrme-type
model as $r \rightarrow \infty$, the
 pion field asymptotes
to a p-wave Yukawa form with a field
 strength fixed by $g_{\pi N N}$:
\begin{equation}
\pi(r) \rightarrow \pi_{\rm asympt}(r) =
 \frac{3 g_{\pi N N}}{8 \pi M_N} \, (m_\pi +
 1/r) \, \frac{{\rm e}^{-m_\pi
 r}}{r}
\end{equation}
 The isovector charge radius is given by
\begin{equation}
\langle r^2 \rangle_{I=1} \, = \,
 \frac{1}{\cal I} \, \int \, {\rm d}r \, r^4
 \, \frac{8 \pi}{3} ({\pi_{\rm asympt}}^2  +
 A(r))
\end{equation}
where $A(r)$ is a model-dependent function
 which goes to zero faster than
${\pi_{\rm asympt}}^2$ and ${\cal I}$ is the
 moment of inertia.  Inserting the asymptotic form into the
integral, using the
 Goldberger-Trieman relation\cite{GT}
 $g_{\pi N N} f_{\pi} = g_{A}  M_N$ (which
 is true to leading order in the pion mass)
 and the fact that $1/{\cal I} = 2/3
 (M_\Delta -M_N)$ (to leading order in the
 1/N expansion) immediately yields
 eq.~(\ref{r2s}) + corrections which are
 higher order in either $1/N_c$ or $m_q$.
 It is worth stressing that this result is
 model independent in the sense that it
 applies to all Skyrme-type models.

Of course, this result is incompatible with
 the known chiral behavior of the charge
 radius in chiral perturbation theory
 \cite{chi1,chi2}: $\langle r^2
 \rangle_{I=1} \, = \, - \frac{5 g_A^2+1}{8
 \pi^2 f_\pi^2} {\rm Log} (m_\pi/\Lambda)$,
 where $\Lambda$ is a mass parameter
 independent of $m_q$.
This was originally deduced using dispersive
 methods\cite{chi1}. In more modern language
 this result can be obtained from
heavy baryon effective chiral lagrangian
 methods \cite{chi3,chi4}. In such
 treatments the leading nonanalytic behavior
 in $m_\pi^2$ is believed to be completely
 reproduced by 1 pion loop feynman graphs.
  The term proportional to $5 g_A^2$ comes
from diagram a. in fig. (\ref{fig}) while
 the term proportional
to unity comes from diagram b.  To the
 extent, that we are interested in the large
 $N_c$
behavior, the term proportional to unity may
 be dropped since $g_A \sim N_c$.

The nonanalytic behavior in $m_\pi$ in
 diagram a. in this standard chiral
 perturbation theory analysis  comes
entirely from the nearly vanishing
energy denominators associated with the
 $\pi$-N states.  However, in the large $N_c$ limit of QCD the
$\Delta$ becomes
 nearly degenerate with the nucleon; the
 mass splitting goes as $1/N_c$.  This can be seen
 directly in the Skyrme model \cite{ANW,sc2}
 and can also be deduced in a model-independent
way by large $N_c$ consistency
 conditions\cite{lnc}.
Thus, in an effective lagrangian treatment
 of the combined large $N_c$ and chiral
 limits, one must treat the $\Delta$ as an
 explicit low energy degree of freedom.
 Indeed, in the large $N_c$ limit there is
 an entire band of low-lying states with
 $I=J$.  As will be shown, however, for the
 leading   nonanalytic behavior in $1/N_c$
 and $m_q$ of nucleon properties, only the N
 and $\Delta$ contribute.

Moreover, in the large $N_c$ limit the
 $\pi$-N-$\Delta$ coupling is completely
 fixed.  It is sufficient to specify the
 coupling in terms of a heavy baryon
 effective field theory (HBEFT)\cite{chi3}
 in which the baryon degrees of freedom are
 treated nonrelativistically.  The HBEFT is
 the natural formulation of chiral
 pertubation in which chiral power counting
 rules can be satisfied.  The effective
 $\pi$-N-$\Delta $ coupling in the large
 $N_c$ limit is given by
\begin{equation}
{\cal L}_{\pi{\rm N} \Delta} \, = \,
\frac{3 g_A}{f_\pi}
 (\overline{\Delta}_{m^\prime, m_I^\prime}\,
 X^{\Delta N \, i a}_{(m^\prime, m_I^\prime)
 (m, m_I)} \, N_{m,m_I} \, \partial_i \pi_a
 \; + {\rm h.c.} )
\label{c1}\end{equation}
where
\begin{equation}
X^{\Delta N \, i a}_{(m^\prime, m_I^\prime)
 (m, m_I) } \, = \, \frac{1}{\sqrt{2}} \,
 \left ( \begin{array}{c c c} 1/2 & 1& 3/2\\
m_I & a & m^\prime_I \end{array}\right) \,
\left ( \begin{array}{c c c} 1/2 & 1& 3/2 \\
m & i & m^\prime \end{array}\right)
\label{c2}\end{equation}
The couplings in eqs. (\ref{c1}) and
 (\ref{c2}) are model independent; they can
 be obtained directly from the Skyrme model
 using conventional semiclassical
 techniques.  This can also be obtained from large $N_c$
consistency relations\cite{lnc}.
Indeed, the large $N_c$ consistency
 relations can be used to show that any
 correction to these couplings is of
 relative order $N_c^{-2}$.

The notion that the $\Delta$ should be
 included as an explicit degree of freedom
in chiral perturbation theory is not novel.
 It has been argued
on phenomenolgical grounds\cite{chi3}  that
 the $\Delta$ (or more generally the
 decuplet in the three flavor case)
should be included since it is light in the
 real ($N_c=3$) world. The formal need to
 include the $\Delta$  explicitly in the
 combined large $N_c$ and chiral limits is
 completely clear and has been seen
 unambiguously in the nonanalytic behavior
 of vector-isovector
and scalar-isoscalar observables \cite{cb}.

The leading order nonanalytic behavior in
the combined large $N_c$ and chiral limits
 (with $N_c m_\pi$ fixed) presumably comes
 from one pion loops with either nucleons or
 Deltas in the intermediate state.  Thus, in
 addition to the graph in diagram (a.) of fig.
 (\ref{fig}) one needs to include diagram (c.).
  Note that other states in the $I=J$ band
 ({\it i.e.} states with $I=J \ge 5/2$) do
 not contribute to the leading nonanalytic
 behavior since, by isospin, at least two
 pions are required to connect to the
 $I=J=5/2$ state and more to higher states
 while the leading nonanalytic behavior comes
 from a single pion loop. Evaluation of this
 loop is straightforward  in HBEFT.  The
 contribution to the isovector charge radius
 from this loop using the large $N_c$ $\pi$-N-$\Delta$ vetex
from eqs. (\ref{c1}) and (\ref{c2}) is
\begin{equation}
\langle| r^2 | \rangle_{I=1}^{\Delta \rm
 loop} \, = \,    \frac{5 g_A^2}{8 \pi^2
 f_\pi^2} {\rm Log} (m_\pi/\Lambda) \,  +
 \frac{5 \, g_A^2 }{4 \, \pi^2 f_\pi^2}
 \frac{\delta m}{\sqrt{m_\pi^2 - \delta
 m^2}} {\rm
 tan}^{-1}\left(\sqrt{\frac{m_{\pi} - \delta
 m}{m_\pi + \delta m}}\right )
\end{equation}
Note that the Log term precisely cancels
the Log  from diagram a.
and the sum of the two immediately gives
eq. (\ref{r2e}).

It is useful to rewrite eq. (\ref{r2e}) in
 terms $d \equiv \delta m/m_{\pi}$:
\begin{equation}
\langle {\rm N}| r^2 |{\rm N} \rangle_{I=1}
 \, = \frac{5 \, g_A^2 d}{16 \pi  f_\pi^2}
 \, S(d) + C
\label{a} \end{equation}
where C is subleading and $S(d)$ is a
 suppression factor associated with the fact
 that the $\Delta$ is not degenerate
with the nucleon:
\begin{equation}
s(d)= \frac{4}{\pi} \, \frac{1}{\sqrt{1
 -d^2}} \, {\rm tan}^{-1}
 \left(\sqrt{\frac{1 - d}{1+d}} \right)
\label{b}\end{equation}
This suppression factor $s(d)$ is identical
 to the one that appears in the leading
 nonanalytic behavior of scalar-isoscalar
 and vector-isovector
as shown in ref. \cite{cb}.

A few observations are in order about eqs.
 (\ref{a}) and (\ref{b}).  The first
is that if one takes the large $N_c$ limit
 with $m_\pi$ held fixed then $d$ goes to
 zero.  Since $s(0) = 1 $, it is easy to see
 that one immediately reproduces the Skyrme
 model result of eq. (\ref{r2s}) in the
 formal $N_c \rightarrow \infty$ limit. This
 is significant in that it supports the
 conjecture that the Skyrme model correctly
 reproduces all of the model-independent
 predictions of large $N_c$ QCD.   A second
 issue is the treatment of eq. (\ref{b})
 when $d > 1$.  This is of immediate
 consequence to the real world in which $d
 \sim 2$.  The basic point is that the
 expression must be analytically continued.
 In principle, there is an ambiguity in this
 continuation since there is a branch cut at
 $d=1$.  However, this ambiguity is
 trivially resolved by the realization that
 the imaginary part must vanish for this
 observable.  Thus for $d > 1$
\begin{equation}
 s(d)= \frac{4}{\pi} \, \frac{1}{\sqrt{d^2
 -1}} \, {\rm tanh}^{-1} \left(\sqrt{\frac{d
 - 1}{d+1}} \right)
\label{c}\end{equation}
The formal limit of $m_{\pi}\rightarrow 0$
 is easy to take.  One obtains $\langle r^2
 \rangle_{I=1} \, = \, - \frac{5 g_A^2}{8
 \pi^2 f_\pi^2} {\rm Log} (m_\pi/\delta m)$.
  This disagrees with the correct chiral
perturbation theory result by an overall
factor of $5 g_A^2/(5 g_A^2 +1)$.  The
 disagreement, however, is of order $N_c^{-2}$
 relative to the leading order and is due to
 the neglect of diagram (b.) of fig.
 (\ref{fig}).

Using eqs.~(\ref{a}) and (\ref{c}) and the
 physical values for
$g_A$, $\delta m$, $f_\pi$ and $m_\pi$ yields
$\langle r^2 \rangle_{I=1} =.72 \, {\rm fm}^2 +
 C$ where C is subleading.  The experimental
 value is .88 ${\rm fm}^2$.  Thus, this
 simple leading order result gives the
 correct result to within about 20\% without
 including effects such as vector dominance.

It is worth stressing that the pure $1/N_c$
expansion (with $m_{\pi}$ fixed) for any
 quantity which depends on $s(d)$ is highly
 problematic.  The  value of  $d$ in the real
 world is approximately 2.1  and
$s(d_{\rm phys}) \approx .47$ .  This is
 quite far from the large $N_c$ value of
 unity.  Ideally, one might hope that the
 disagreement between $s(0)$ and $s(d_{\rm
 phys})$  can be explained in terms of
 higher order terms in the $1/N_c$ expansion
 which is just a power series in $d$ around
 $d=0$. Unfortunately the radius convergence  of this series is
unity.  Thus,  the
 physical value of
$d \approx 2.1$ is beyond the radius of
convergence.  This implies that the inclusion
 of any  finite number of subleading terms
 in the $1/N_c$ expansion will not improve
 the description and will, in general make
 things worse: denoting the Taylor expansion
 of $s(d)$ up to $n^{\rm th}$ order by
 $s^n(d)$ one finds, for example,  $s^0(d_{\rm
 phys})=1$, $s^1(d_{\rm phys})\approx -.34$,
 $s^2(d_{\rm phys}) \approx 1.8$ and
 $s^3(d_{\rm phys}) \approx -2.06$.  The
 obvious lesson from this is that the pure
 $1/N_c$ is not likely to be convergent for
 the isovector charge radius and is
 certainly not convergent if it is dominated
 by pionic tail effects.  This problem is
 particularly serious for Skyrme-type
 models---the only systematic calculational
 scheme currently known for these models is
 the pure $1/N_c$ expansion.

It is clear that analogous effects will
follow for any isovector electric property,
 such as higher moments of the electric form
 factor or moments of the N-$\Delta$
 transition form factor. The Skyrme model
 will correctly reproduce the leading order
 large $N_c$ and chiral behavior provided
 that the large $N_c$ limit is taken first.
  However, the corrections to
this leading order $1/N_c$ behavior for
 physical parameters will be large and the
$1/N_c$ expansion will not converge.

  The distinction between the leading order
 large $N_c$ results and the physical
 results is particularly important in the
 case of properties of the $\Delta$
or of the N-$\Delta$ transition.  The
 difficulty is that for $N_c \rightarrow
 \infty$ the $\Delta$ is stable and
 quanitities such as the $\Delta$ charge
 radius or quadrupole moment have an
 unambiguous meaning in terms of measureable
 quantities.  However, as $N_c$ decreases to
 the point that the $\Delta$ becomes
 unstable these quanitities become ambiguous
 in the sense that there is no simple
model-independent connection between the
 calculated quanitity and experimental
 observables.

An important example of this is the
 N-$\Delta$ electric quadrupole transition
 matrix element.  In the large $N_c$ and
chiral limits this matrix element with the
 large $N_c$ limit taken first, the matrix
 element corresponds to a well-defined
 observable directly proportional to the
 isovector charge radius.  A quasi-model
 independent  relationship between the two
 \cite{sc2} becomes exact in the chiral
 limit.  This gives a transition quadrupole
 matrix element which is much
larger than seen in typical quark models.
 However, as one increases $d$ beyond unity
  the matrix  element ceases
to have any well-defined experimental signature.  Thus, it is
hard to use the quadrupole transition matrix element to
distinguish between various models of the baryon.

The author gratefully acknowledges the
 hospitality of the Institute for Nuclear
 Theory and the Physics Dapartment of the
 University of Washington.  This work was
 supported by the U. S. Department of Energy
 (grant \#DE-FG02-93ER-40762) and National Science Foundation (grant
\#PHY-9058487).

\begin{figure}
{\noindent \bf Fig. 1: \rm Feynman graphs for
leading chiral behavior of  $\langle r^2 \rangle_{I=1}$.
 The cross represents an insertion of the $0^{\rm th}$
component of the electromagnetic current, the dotted
line is a pion propagator, the solid line  is a
nucleon propagator and the double line is
 the $\Delta$ propagator.}
\label{fig}

\end{figure}

\begin{references}
\bibitem{sc} This quantization scheme was
introduced in  \cite{ANW}.
 A reformulation of this method, valid for
models which include fermions or vector
 particles may be found in \cite{sc2}.
\end{references}
\end{document}